\documentclass[11pt]{article}
\usepackage{graphicx}

\begin{document}

\def\xslash#1{{\rlap{$#1$}/}}
\def \p {\partial}
\def \dd {\psi_{u\bar dg}}
\def \ddp {\psi_{u\bar dgg}}
\def \pq {\psi_{u\bar d\bar uu}}
\def \jpsi {J/\psi}
\def \psip {\psi^\prime}
\def \to {\rightarrow}
\def\bfsig{\mbox{\boldmath$\sigma$}}
\def\DT{\mbox{\boldmath$\Delta_T $}}
\def\xit{\mbox{\boldmath$\xi_\perp $}}
\def \jpsi {J/\psi}
\def\bfej{\mbox{\boldmath$\varepsilon$}}
\def \t {\tilde}
\def\epn {\varepsilon}
\def \up {\uparrow}
\def \dn {\downarrow}
\def \da {\dagger}
\def \pn3 {\phi_{u\bar d g}}

\def \p4n {\phi_{u\bar d gg}}

\def \bx {\bar x}
\def \by {\bar y}

\begin{center}
{\Large\bf  Single Transverse Spin Asymmetries at Parton Level }
\par\vskip20pt
H.G. Cao$^1$, J.P. Ma$^2$  and H.Z. Sang$^{2,3}$    \\
{\small {\it  $^1$ Science School, Beijing University of Civil Engineering and Architecture, Beijing 100044, China
\\
$^2$ Institute of Theoretical Physics, Academia Sinica,
P.O. Box 2735,
Beijing 100190, China\\
$^3$ Institute of Modern Physics,
School of Science,
East China University of Science and Technology,
130 Meilong Road, Shanghai 200237, P.R. China
}} \\
\end{center}
\vskip 10mm
\begin{abstract}
Two factorization approaches have been proposed for single transverse spin asymmetries.
One is the collinear factorization, another is the transverse-momentum-dependent
factorization.
They have been previously derived in a formal way
by using diagram expansion at hadron level. If the two
factorizations hold or can be proven, they should also
hold when we replace hadrons with parton states.
We examine these two factorizations at parton level
with massless partons. It is nontrivial to generate these asymmetries at parton level with massless
partons because the asymmetries require helicity-flip and nonzero  absorptive parts in scattering amplitudes.
By constructing suitable parton states with massless partons
we derive the two factorizations
for the asymmetry in Drell-Yan processes. It is found from our results that  the
collinear factorization derived at parton level is not the same
as  that derived at hadron level.
Our results with massless partons confirm those derived
with single massive parton state in our previous works.

\end{abstract}
\vskip 1cm

\par
\noindent
\par\noindent
{\bf 1. Introduction}
\par
Single transverse-spin asymmetries(SSA) have been observed in various experiments.
An updated
review about the phenomenology of SSA can be found in \cite{Review}.
These asymmetries are expected if scattering amplitudes have
nonzero absorptive parts and helicity-flip interactions are involved.
The observed
asymmetries provide a new tool to study the structure because they are sensitive
to the correlations of partons inside a hadron and the angular momenta of these partons.
In general it is difficult to predict the size of observed SSA
in a process involved with hadrons, especially,
with a transversely polarized hadron,
because we have not enough information about the structure of hadrons.
However, for SSA observed in processes involving large momentum transfers certain predictions
can be made by using the concept of QCD factorization. In the factorization
perturbative- and nonperturbative effects are separated. The nonperturbative effects
are represented by various matrix elements of QCD operators and these matrix elements contain
information about the structure of hadrons.
\par
Two factorization approaches have been proposed. One is the collinear factorization
in which the partons in a hadron entering hard scattering only carry the momenta
parallel or collinear to the hadron. Another is
the transverse-momentum-dependent(TMD) factorization, where one takes transverse momenta of partons in
hadrons into account.
It should be noted that in the two approaches the factorization is derived or proposed rather formally in
the sense that one works at hadron level by using the diagram expansion.
The purpose of our work is to examine the two factorization approaches for SSA in Drell-Yan
processes and Semi Inclusive DIS(SIDIS)
at parton level with massless partons.
\par
The motivation of our work is based on the following: QCD
factorizations, if they are proven, are general properties of QCD.
These factorizations hold not only with hadron states but also hold when one replaces
the hadron states with parton states. This is in the sense that
the perturbatively calculable parts in factorizations do not depend
on hadrons and  are completely determined by hard scattering of
partons. In other words the perturbatively calculable parts or hard parts can be
extracted if we replace hadronic states with partonic states to
calculate ingredients in factorizations. This fact can be
illustrated by the typical example of the factorization in DIS,
where one can use a parton state to calculate the structure
functions and parton distribution functions to extract the hard
part. Because at the leading twist of the factorization only one parton is struck, a
parton state with a single parton is enough for doing this.
QCD factorizations can
also be understood from the point of view of effective theories
where a factorization is understood as a matching procedure of
operators of full QCD into operators of an effective theory. The
matching coefficients are the perturbatively calculable parts.
The matching procedure is done with parton states,  where one requires matrix elements
of full QCD operators with partonic states to be the same of operators of an effective theory
with the same partonic states multiplied with matching coefficients.
\par
There are distinct differences between two factorization approaches for SSA.
In the approach of collinear factorization, the nonperturbative effects responsible for SSA
are represented  by twist-3 matrix elements\cite{QiuSt, EFTE,KaKo,EKT},
or called ETQS matrix elements. These twist-3 matrix elements only contain the effect
of spin-flip interactions. The nonzero absorptive part or $T$-odd effect
is generated by poles of parton propagators in hard scattering.
Applications of the collinear factorization for SSA can be found in \cite{tw31,tw32}.
In the approach of TMD factorization, the nonperturbative effects responsible for SSA
are represented by matrix elements containing $T$-odd- and spin-flip effects. These matrix
elements are the so-called Sivers\cite{Sivers}- and Collins\cite{JC} functions.
So far TMD factorization has been studied carefully in \cite{CS,CSS,JMY,JMYG,CAM} only for physical quantities
which do not contain $T$-odd effects. In these studies
the factorization is examined by replacing hadrons with partons and it shows that
TMD parton distributions
entering the factorization can be defined with QCD operators
consistently. For SSA as $T$-odd effect intensive efforts has been spent
to study how to consistently define or interpret Sivers function as a parton distribution
which is gauge invariant and contains initial- or final state interactions\cite{JC,SJ1,TMDJi,Mulders97,Boer03}.
Through these studies consistent definitions of these distributions or fragmentation functions which
contain $T$-odd effects can be given. With Sivers functions SSA has been studied extensively
\cite{Anselmino,Mulders,DeSanctis,Efremov,BQMa}.
The TMD factorization is expected to hold at low transverse momentum region.
It is interesting to note that for SIDIS and Drell-Yan both approaches are applicable
for certain kinematic region and it can be shown that the two approaches are equivalent\cite{JQVY1,JQVY2,KVY}.
\par
The two factorization approaches for SSA are derived rather formally in
the sense that one works with a diagram expansion involving hadrons,
where one separates diagrams involving hadrons
into a parton-scattering part and hadronic parts.  The hadronic parts
are characterized by various parton density matrices of hadrons.
With the expansion the perturbative coefficients
in the two factorizations are then determined at the leading order of $\alpha_s$.
As discussed in the above, the two factorizations should
also hold when hadrons are replaced with partons,
i.e., the perturbative coefficients extracted with partonic states should be the same
as those derived with the diagram expansion at hadron level. It should be emphasized
for a correct collinear factorization of SSA that the relevant structure function
calculated with a parton state should be reproduced from the factorization formula
by using the twist-3 matrix element calculated with the same parton state, even if the parton state
may be simple so that one can not derive the complete hard part in the factorization.
This is because the hard part should not depend on hadrons and hence it should also not depend
on the parton states which is used to replace the hadrons.
The same also holds for the TMD factorization.
This will be discussed in detail when we give all results.
\par
We have examined the two approaches of factorization with partonic states
for SSA in Drell-Yan processes\cite{MS1} and SIDIS\cite{MS2}, where we have replaced
the transversely polarized hadron with a quark. In order to have helicity-flip effect at parton level,
the quark can not be taken as massless. We give the quark a small mass $m$ and calculate
every element in the two factorizations to examine them. Although the calculation
is done at leading order, but it is nontrivial because absorptive parts of amplitudes
are needed.
It is well known that SSA involved with a massive quark is proportional to the quark mass $m$
\cite{KPR, DGBB}. SSA's calculated in \cite{MS1,MS2} are also proportional to $m$.
The derived perturbative coefficients in the factorization formulas do not depend on the quark mass $m$,
because the twist-3 matrix element and the Sivers function in the corresponding
factorization are also proportional to the quark mass $m$.
Our results show that the formally derived TMD factorization for SSA holds at parton level,
while the perturbative coefficient extracted from partonic results in
the collinear factorization is different than that in the formally derived collinear factorization.
In other word, the SSA calculated with the parton state can not be reproduced with the formally derived
collinear factorization by using the twist-3 matrix element calculated with the same
parton state.
The collinear factorization can be derived with the massive parton state
at leading but nontrivial order of $\alpha_s$, but
the leading order result of the twist-3 matrix element
responsible for SSA is U.V. divergent and calls for renormalization. The collinear
factorization is then derived by interpreting the renormalization scale $\mu$
as an effective cutoff for the observed transverse momentum. This may be unsatisfied and
may bring some problems at higher orders of $\alpha_s$.
\par
The purpose of the current work is to study the two factorizations with massless partons.
We construct a suitable parton state of two components  to replace the transversely
polarized hadron. One component consists of one quark,
another consists of one quark and one gluon. With this parton state one can obtain nonzero SSA and
calculate nonperturbative matrix elements for extracting perturbative coefficient
functions in the two factorizations.
With massless partons we confirm all factorization results in our
previous works\cite{MS1,MS2} where a single massive quark for the replacement.
We establish a modified  relation between the twist-3 matrix element
and the first $k_\perp$-moment of Sivers function. The relation has been derived
before in \cite{Boer03,MW1}, where the problem of U.V. subtractions has not been considered.
\par
Our work is organized as the following: In Sect.2 we
construct a partonic state which is suitable to study SSA at parton level with massless partons.
Then we derive the leading order result of the twist-3 matrix element with the partonic state.
This result shows how the correlation between spins of partons generates helicity-flip effect
even if the partons are massless.
In Sect. 3 we give detailed results for SSA in Drell-Yan processes with the parton states
and derive the collinear factorization, which is the same as derived in \cite{MS1} with massive partons.
In Sect. 4 we study Sivers function with the parton state and derive TMD factorization with our partonic
results. In Sect.5 the relation between the twist-3 matrix element and the first $k_\perp$-moment
of Sivers function is addressed. Sect.6 is our summary.

\par\vskip20pt
\noindent
{\bf 2.  Partonic State and Twist-3 Matrix Element}
\par
Before we construct the suitable partonic state with massless partons to study SSA,
we recall some standard descriptions
of spin-1/2 systems from text books.
We consider a system $\vert n[\lambda]\rangle$ with the total spin $1/2$.
The system can consist of more than one partons.
Its helicity is given by $\lambda =\pm$.
Now we consider a forward scattering of the system through some operator ${\mathcal O}$. The
transition amplitude is given as:
\begin{equation}
 {\mathcal M}_{\lambda_2 \lambda_1} = \langle  n[\lambda_2 ] \vert {\mathcal O} \vert n [\lambda_1 ]\rangle.
\end{equation}
The system moves in the $z$-direction. We use $\lambda_{1,2}=\pm $ to denote the helicity of the
initial- and final state, respectively. The transition amplitude in the helicity space is
$2\times 2$ matrix and can be expanded as:
\begin{equation}
{\mathcal M}_{\lambda_2 \lambda_1} = \left [ a + \vec b \cdot \vec \sigma \ \right ]_{\lambda_2 \lambda_1 }
\end{equation}
with $\sigma^i (i=1,2,3)$ are Pauli matrices. The spin of the system can also be described by
a spin vector $s^\mu =(s^0, \vec s)$ with $s^2=-1$.
The two descriptions are equivalent because the transition
amplitude characterized by the vector $s$ can be written with the same coefficient
$a$ and the same vector $\vec b$ in Eq.(2):
\begin{equation}
{\mathcal M} (s)  = \left [ a + \vec b \cdot \vec s \  \right ].
\end{equation}
Single spin azimuthal asymmetries can in general appear if the non-diagonal part of ${\mathcal M}$ of
certain operators in the helicity
space, i.e., ${\mathcal M}_{+-, -+}$ are nonzero, or, the transverse part of the vector
$\vec b$, i.e., $b^1$ and $b^2$, nonzero.
\par
In this work we will use the  light-cone coordinate system, in which a
vector $a^\mu$ is expressed as $a^\mu = (a^+, a^-, \vec a_\perp) =
((a^0+a^3)/\sqrt{2}, (a^0-a^3)/\sqrt{2}, a^1, a^2)$ and $a_\perp^2
=(a^1)^2+(a^2)^2$. In the light-cone coordinate system we introduce
two light-cone vectors: $n^\mu=(0,1,0,0)$ and $l^\mu=(1,0,0,0)$. We define the totally
antisymmetric tensor in the transverse space as $\epsilon_\perp^{\mu\nu} =\epsilon^{\alpha\beta\mu\nu}l_\alpha n_\beta$
with $\epsilon^{0123}=1$.
\par
We consider the partonic state with massless partons:
\begin{equation}
 \vert P [\lambda ] \rangle  =  \vert q(p,\lambda_q) [\lambda ] \rangle + c_1
                   \vert q(p_1,\lambda_q) g(k,\lambda_g ) [\lambda ] \rangle,
\label{state}
\end{equation}
with the momentum specified as:
\begin{equation}
   p^\mu = P^\mu = (P^+,0,0,0), \ \ \  p_1^\mu = x_0 P^\mu, \ \ \ \ k^\mu =(1-x_0) p^\mu =\bar x_0 p^\mu.
\end{equation}
The state of the system is a superposition of a single quark state and a quark-gluon state.
Because we will replace for our purpose an initial hadron with the partonic state, all partons in the state
is in the in-state.
The state has the helicity $\lambda$ and the total momentum $P$. The single quark state has the same helicity
$\lambda_q =\lambda$.
For the $qg$-state, the total helicity is the sum $\lambda=\lambda_q + \lambda_g $.
The $q$-state and $qg$-state carries the same color index $i_c$ as given by
\begin{equation}
\vert  q (p,\lambda_q) \rangle = b^\dagger_{i_c} (p,\lambda_q) \vert 0 \rangle,
\ \ \ \ \
\vert q (p_1,\lambda_q)g(k,\lambda_g) \rangle =
T^a_{j_c i_c} b^\dagger_{j_c} (p_1,\lambda_q) a^\dagger_a (k,\lambda_g) \vert 0 \rangle,
\end{equation}
where $b^\dagger_i$ is the quark creation operator with $i$ as the color index,
$a^\dagger_a$ is the gluon creation operator with $a$ as the color index.
$c_1$ is taken as a real number. We take $c_1=1$ for simplicity.
Since the quark is massless, the non-diagonal part of the transition amplitude in Eq.(1) calculated in
perturbative theory of QCD will obtain contributions only from
the interference between the single quark- and the quark-gluon state:
\begin{eqnarray}
{\mathcal M}_{+ - } &=& \langle q(p,+) [ + ] \vert {\mathcal O} \vert q(p_1,+ ) g(k, - ) [- ]  \rangle
                        + \langle q(p_1,- ) g(k, +  ) [ + ] \vert {\mathcal O} \vert q(p,-) [-] \rangle,
\nonumber\\
{\mathcal M}_{- + } &=& \langle q(p,-) [ - ] \vert {\mathcal O} \vert q(p_1,- ) g(k, + ) [ + ]  \rangle
                        + \langle q(p_1,+ )g(k,-) [ - ] \vert {\mathcal O} \vert q(p,+) [+] \rangle.
\end{eqnarray}
In the above the helicity of the quark is not flipped and the color
of the state is summed, i.e., the color index $i_c$ given in Eq.(6) is summed in Eq.(7). It is clear
from Eq.(7) that for SSA one needs to study the forward scattering of the $qg$-state into the $q$-state
and the $q$-state into the $qg$-state.
\par
Now we consider the twist-3 matrix element relevant to SSA. The definition with a transversely polarized
hadron state is:
\begin{eqnarray}
T_F (x_1,x_2)  \epsilon_\perp^{\mu\nu}
s_{\perp\nu}
   & =&    \frac{g_s}{2}\int \frac{dy_1 dy_2}{4\pi}
   e^{ -iy_2 (x_2-x_1) P^+ -i y_1 x_1 P^+ }
\nonumber\\
    && \cdot \left  \{ \langle P, \vec s_\perp \vert
           \bar\psi (y_1n ) \gamma^+ G^{+\mu}(y_2n) \psi(0) \vert P,\vec s_\perp \rangle
  - ( \vec s_\perp \to - \vec s_\perp ) \right \}.
\label{tw3}
\end{eqnarray}
In the above we have suppressed the gauge links along direction $n$ between operators. These gauge links
make the definition gauge invariant.
If we replace the hadronic state with the state $\vert P [\lambda ]\rangle$
we can calculate $T_F$ perturbatively.
With the definition we can find the operator for the transition in Eq.(1) as:
\begin{equation}
{\mathcal O}^\mu  = g_s\int \frac{dy_1 dy_2}{4\pi}
   e^{ -iy_2 (x_2-x_1) P^+ -i y_1 x_1 P^+ }
           \bar\psi (y_1n ) \gamma^+ G^{+\mu}(y_2n) \psi(0).
\end{equation}
It is straightforward to calculate the non-diagonal part of the transition amplitude
at leading order of $\alpha_s$. We have:
\begin{eqnarray}
{\mathcal M}^\mu_{+ - } &=&  ig_s\pi \sqrt{x_0} (N_c^2-1) (x_2-x_1) \left [ \delta (1-x_1) \delta (x_2-x_0)\epsilon^\mu(-)
        +  \delta (1-x_2) \delta (x_1-x_0) \epsilon^{* \mu}(+) \right ],
\nonumber\\
{\mathcal M}^\mu_{- + } &=&  ig_s\pi \sqrt{x_0} (N_c^2-1)(x_2-x_1) \left [ \delta (1-x_1) \delta (x_2-x_0)\epsilon^\mu(+ )
        +  \delta (1-x_2) \delta (x_1-x_0) \epsilon^{* \mu}(-) \right ],
\end{eqnarray}
where $\epsilon^\mu(\pm)$ is the polarization vector of the gluon.
Using the representation of the polarization vector of the gluon moving in the $z$-direction:
\begin{eqnarray}
 \epsilon^\mu (+) = -\frac{1}{\sqrt{2}} (0,0,1,i), \ \ \ \   \epsilon^\mu (-) = \frac{1}{\sqrt{2}} (0,0,1,-i),
\ \ \ \ \epsilon^{*\mu} (+) = -\epsilon^\mu (-),
\end{eqnarray}
we can express the non-diagonal part with $\vec \sigma_\perp =(\sigma^1, \sigma^2)$ in a matrix form in
the $2\times 2$ helicity space:
\begin{eqnarray}
{\mathcal M}^\mu =  -g_s\pi \sqrt{\frac{x_0}{2}} (N_c^2-1)(x_2-x_1) \epsilon^{\mu i}_\perp \sigma^i_\perp  \left [
                      \delta (1-x_1) \delta (x_2-x_0) - \delta (1-x_2) \delta (x_1-x_0) \right ],
\end{eqnarray}
Therefore we have the nonzero twist-3 matrix element of our partonic state at leading order of $\alpha_s$:
\begin{equation}
T_F(x_1,x_2)= g_s\pi \sqrt{\frac{x_0}{2}} (N_c^2-1)(x_2-x_1)  \left [
                      \delta (1-x_1) \delta (x_2-x_0) - \delta (1-x_2) \delta (x_1-x_0) \right ].
\end{equation}
This gives an example showing how the effect of helicity flip is generated by the correlation
between the spin of a massless quark and that of a gluon.
We note here at leading order $T_F(x,x)=0$ and also that $T_F(x_1,x_2)/(x_2-x_1)$ is approaching to zero
with $x_2$ approaching $x_1$. It should be noted that there is no proof in QCD to show $T_F(x,x)=0 $ or
$T_F(x,x)\neq 0$. With our partonic state we have $T_F(x,x)=0$ at tree-level. At one-loop level
$T_F(x,x)$ becomes nonzero, as we will show.
Another interesting quantity which will enters the collinear factorization for SSA is defined as:
\begin{eqnarray}
\tilde T_F (x_1,x_2)
s_{\perp}^{\mu}
   & =&    -i\frac{g_s}{2}\int \frac{dy_1 dy_2}{4\pi}
   e^{ -iy_2 (x_2-x_1) P^+ -i y_1 x_1 P^+ }
\nonumber\\
    && \cdot \left  \{ \langle P, \vec s_\perp \vert
           \bar\psi (y_1n ) \gamma^+ \gamma_5 G^{+\mu}(y_2n) \psi(0) \vert P,\vec s_\perp \rangle
  - ( \vec s_\perp \to - \vec s_\perp ) \right \}.
\label{tw31}
\end{eqnarray}
Replacing the hadron state with the partonic state in Eq.(4) we have the leading order result:
\begin{equation}
\tilde T_F(x_1,x_2)= g_s\pi \sqrt{\frac{x_0}{2}} (N_c^2-1)(x_2-x_1)  \left [
                      \delta (1-x_1) \delta (x_2-x_0) + \delta (1-x_2) \delta (x_1-x_0) \right ].
\end{equation}

\par
It should be noted that our analysis in this work can be done by constructing a general partonic state.
With such a general parton state the analysis
has some similarities to these works in \cite{BJY,BDH} in different perspectives. In \cite{BJY} a hadron is
represented by its various light-cone components with hadronic wave functions. With the representation
the standard Parton Distribution Function(PDF) is calculated in terms of these wave functions for the
purpose to study the relation between scale evolutions of wave functions and of PDF's. If one uses
the same representation to calculate the standard DIS,
one can extract the perturbative coefficient function of the factorization of the standard DIS.
In \cite{BDH} one represents a hadron with its various light-cone components associated with
hadronic wave functions to study nonforward PDF's in deeply virtual Compton Scattering(DVCS).
The nonforward PDF's are then represented in terms of hadronic wave functions.
Again, one can also use the representation to calculate DVCS for extracting the perturbative
coefficient function in the factorization of DVCS. In both cases one will find that
the extracted perturbative coefficient functions are the same as those extracted
with a single parton state, because in these two cases only two partons are active or entering
hard scattering. Slight differences exist between these works of \cite{BJY,BDH} and ours.
In \cite{BJY,BDH} the wave functions are those of a real hadron, while wave functions
in the general parton state
are arbitrary. But this makes no difference in extracting perturbative coefficient functions.
The wave functions in \cite{BJY,BDH} have $k_\perp$-dependence, i.e., the partons also carry
transverse momenta. But these transverse momenta should be much smaller than the scale
in the hard scattering and should be neglected in the hard scattering for a factorization
at leading twist. This also implies that there will be no difference to set $k_\perp=0$ before or
after a relevant calculation to extract perturbative coefficient functions.
\par
Our study with the parton state in Eq.(4) has also certain similarity to the one-loop study
of the structure function $g_2$ or $g_T$ of DIS in \cite{JiOs,BJJL,BKM}. In these studies there is an interesting
problem of how to identify operators to define nonperturbative correlation functions. At tree-level
it is straightforward to obtain a factorized form for $g_2$ in terms of a quark-quark correlation function
which contains one bad component of the quark field.
However, this factorized form can not be consistent because the bad components of quark fields are
not dynamically independent. The bad components can be eliminated through the equation of motion
and the factorized form can be re-expressed in terms of field correlation functions only with good components.
In this way, $g_2$ is actually factorized in terms of a quark-gluon correlation function, which is similar
to $T_F$ in Eq.(8). In our case such a problem does not exist because all field correlation functions for SSA
do not contain any bad component of fields. After identifying the correct operators for $g_2$
one can then study the one-loop correlation of $g_2$ where one essentially
has to study the forward Compton scattering of one parton to one parton, two to one  and one to two partons.
This is similar as we take the simple parton state in Eq.(4) to study factorizations of SSA, indicated by Eq.(7).

\par
\vskip20pt
\noindent
{\bf 3. SSA in Drell-Yan Processes and Its Collinear Factorization}
\par
We consider the Drell-Yan process:
\begin{equation}
  h_A ( P_A, s_\perp) + h_B(P_B) \to \gamma^* (q) +X \to  \ell^-  + \ell ^+  + X,
\end{equation}
where $h_A$ is a spin-1/2 polarized hadron with the transverse spin-vector $s_\perp$.
We take a light-cone coordinate system in which the momenta and the spin are :
\begin{equation}
P_{A,B}^\mu = (P_{A,B}^+, P_{A, B}^-, 0,0),  \ \ \ \  s_\perp^\mu =(0,0, \vec s_\perp).
\end{equation}
$h_A$ moves in the $z$-direction, i.e., $P_A^+$ is the large component. The spin of $h_B$
is not observed. The invariant mass of the observed lepton pair is $Q^2 =q^2$.
The relevant hadronic tensor is defined as:
\begin{equation}
W^{\mu\nu}  = \sum_X \int \frac{d^4 x}{(2\pi)^4} e^{iq \cdot x} \langle h_A (P_A, s_\perp) h_B(P_B)  \vert
    \bar q(0) \gamma^\nu q(0) \vert X\rangle \langle X \vert \bar q(x) \gamma^\mu q(x) \vert
     h_B(P_B)h_A (P_A, s_\perp)  \rangle,
\end{equation}
and the differential cross-section is determined by the hadronic tensor as:
\begin{equation}
\frac{ d\sigma }{ dQ^2 d^2 q_\perp d q^+ d q^- } = \frac{4\pi \alpha_{em}^2 Q_q^2}{3 S Q^2}
    \delta (q^2 -Q^2)
    \left ( \frac {q_\mu q_\nu} {q^2} - g_{\mu\nu} \right ) W^{\mu\nu}, \ \ \  S=2P_A^+ P_B^-.
\end{equation}
\par
We are interested in the kinematical region where $q_\perp^2 \ll Q^2$. The hadronic tensor
at leading twist accuracy has the structure:
\begin{eqnarray}
W^{\mu\nu} &=& - g_\perp^{\mu\nu} W_U^{(1)} + \left ( g_\perp^{\mu\nu}
-2 \frac{q_\perp^\mu q_\perp^\nu} {q_\perp^2}  \right )
   W_U^{(2)}
\nonumber\\
&&  - g_\perp^{\mu\nu} \epsilon_\perp^{\alpha \beta} s_{\perp\alpha} q_{\perp\beta} W_T^{(1)}
 + \left ( s_{\perp\alpha} \epsilon_\perp^{\alpha\mu} q_\perp^\nu
          +s_{\perp\alpha} \epsilon_\perp^{\alpha\nu} q_\perp^\mu -g_\perp^{\mu\nu}
             \epsilon_\perp^{\alpha \beta} s_{\perp\alpha}  q_{\perp\beta} \right ) W_T^{(2)}
\nonumber\\
  &&  +  q_{\perp\alpha} \left ( \epsilon_\perp^{\alpha\mu}  q_{\perp}^\nu
 + \epsilon_\perp^{\alpha\nu} q_{\perp}^\mu \right ) {\vec  q}_\perp \cdot \vec s_\perp W_T^{(3)}
 +\cdots
\end{eqnarray}
In the above, we only give the tensor structures symmetric in $\mu\nu$. $W_T^{(i)}(i=1,2,3)$ represent
$T$-odd effect related to the spin. $W_U^{(1,2)}$ are responsible for unpolarized cross-sections.
$W_T^{(1)}$ contributes to SSA in the region $q^2_\perp \ll Q^2$ which we will study.
All structure functions depend on the kinematical variables $x= q^+/P_A^+$, $y=q^-/P_B^-$ and
$q_\perp$.
\par
To study the collinear factorization of $W_T^{(1)}$ we replace the hadron
$h_A$ with the state $\vert P [\lambda ]\rangle$ in Eq. (\ref{state})
and the hadron $h_B$ with an antiquark $\bar q$
with the momentum $\bar p^\mu =(0,\bar p^-,0,0)$. The spin of $\bar q$ is averaged.
Then we calculate $W_T^{(1)}$ and try to factorize it as a convolution of
$T_F$, the antiquark distribution $\bar q (z)$ and a perturbative coefficient function ${\mathcal H}$.
By the definition of $W_T^{(1)}$ one can identify the operator
for the transition amplitude in Eq.(1).
In order to have a nonzero $q_\perp$ the intermediate state $X$ must contain at least one gluon.
To generate SSA one needs to have a nonzero absorptive part. At leading order of $\alpha_s$
a nonzero absorptive can only arises in the amplitude $P +\bar q \to\gamma^* + g$.
The absorptive part of the amplitude $P +\bar q \to\gamma^* + g$
can be represented by two sets of cut diagrams, which are given in Fig.1 and Fig.3.
A nonzero SSA is obtained through the interference of the absorptive part of the amplitude
$P +\bar q \to\gamma^* + g$ with the tree level amplitude of $q + \bar q \to \gamma^* + g$, the later
is given by the diagrams in Fig.2.
\par
\begin{figure}[hbt]
\begin{center}
\includegraphics[width=12cm]{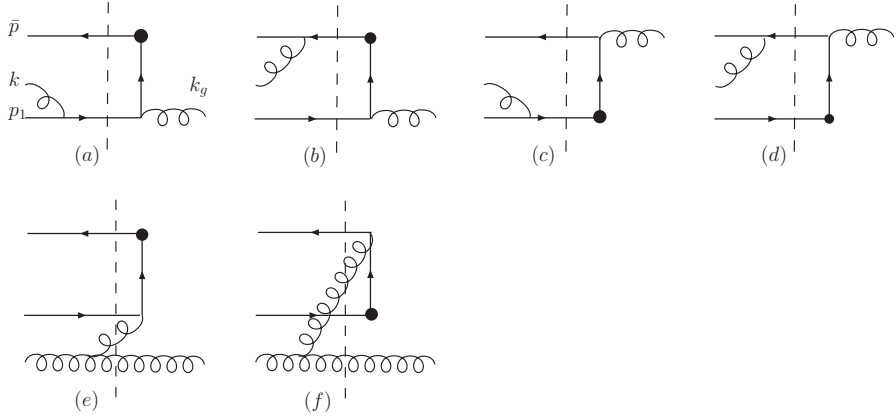}
\end{center}
\caption{Cut diagrams for the absorptive part of the amplitude
$P +\bar q \to \gamma^* + g$. The black dot is for the insertion of the electromagnetic
current, the broken line is for the cut.} \label{Feynman-dg1}
\end{figure}
\par
At first look some diagrams in Fig.1 are dangerous, because in these diagrams like Fig.1a and Fig.1c
one particle absorbs a collinear particle and it will give divergent contributions in the massless limit.
We note that the diagrams
in Fig.1 with or without cuts must exist because of gauge invariance.
It is easy to find that the nonzero absorptive part from cut diagrams in Fig.1 only arise
if the initial gluon has zero momentum, i.e., $k^+=0$.
However, a massless particle state can not be defined with zero momentum. In our case, a zero-momentum
gluon can not be distinguished from the quark, the quark combined with the
zero-momentum gluon should be taken as one quark state. We define our partonic state in Eq.(\ref{state})
with $k^+\neq 0$.
Therefore, we do not need to consider the absorptive parts from Fig.1.
However, it is interesting to explore in more detail why there are no absorptive parts from Fig.1..

\par
\begin{figure}[hbt]
\begin{center}
\includegraphics[width=8cm]{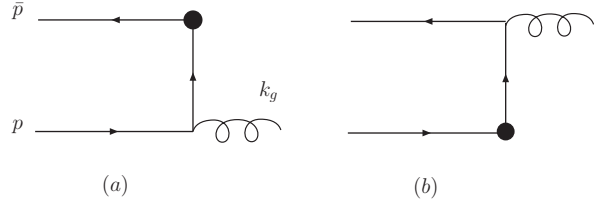}
\end{center}
\caption{Diagrams for the amplitude
$q +\bar q \to \gamma^* + g$. The black dot is for the insertion of the electromagnetic
current. }
\label{Feynman-dg2}
\end{figure}
\par
Taking Fig.1b as an example,
the absorptive part is proportional to
\begin{equation}
\bar v(\bar p)  \gamma \cdot \epsilon(\lambda_g) {\rm Abs}
\left [\frac{\gamma\cdot(-\bar p -k)}{(-\bar p -k)^2+i\varepsilon} \right ]
= -i \frac{\pi}{\bar p^-} \delta (k^+)\bar v(\bar p)\gamma\cdot  \epsilon(\lambda_g)
  \gamma\cdot(-\bar p -k).
\end{equation}
Using the polarization vector given in Eq.(11) and $\bar v(\bar p) \gamma \cdot \bar p =0$, one finds
that the above is proportional to $k^+ \delta (k^+)$. Hence the absorptive part of Fig.1b is zero. For Fig.1a, there is some problem because one always has
$(p_1+k)^2=0$. However, we can give to the quark a small mass and calculate the absorptive part first.
With the massive quark, we have $p_1^\mu =(p_1^+, p_1^-,0,0)$ and $(p_1+k)^2=2 k^+ p_1^- +m^2 $ .
The absorptive part is then proportional to
\begin{equation}
{\rm Abs} \left [ \frac{ \gamma\cdot (p_1 +k) + m} {(p_1+k)^2-m^2+i\varepsilon} \right ]
\gamma \cdot \epsilon(\lambda_g) u(p_1)
=-i \frac{\pi}{p^-_1} \delta (k^+) (\gamma\cdot (p_1 +k) + m )\gamma \cdot \epsilon(\lambda_g) u(p_1).
\end{equation}
By using $  (\gamma\cdot p_1 -m ) u(p_1)=0$ one finds again that the absorptive part of Fig.1a is proportional
to $k^+ \delta (k^+)$ and hence it is zero.
One can also consider the case that the massless quark and gluon have a relative transverse momentum
$k_\perp$. The total transverse momentum of the quark and gluon is zero. In this case one easily
finds that the cut diagram Fig. 1a gives no contribution.
Similarly, one finds
that the absorptive part from Fig.1c and Fig.1d are zero.
However, we find that the absorptive part from Fig.1e and Fig.1f are nonzero with $k^+ =0$.
But as discussed before, we should take the partonic state with $k^+ \neq 0$. Hence the absorptive part
from Fig.1 is always zero.
In the collinear factorization derived formally, $W_T^{(1)}$ receives a piece of contributions
as a convolution with $T_F(x,x)$, called as soft-pole contribution, and a piece of contributions
as a convolution of $T_F(x_1,x_2)$, called hard-pole contribution. The soft pole contributions
at parton level at the order we consider correspond actually to the contributions from Fig.1. From
the above discussion and also the result $T_F(x,x)=0$ in the last section
we can conclude that the soft pole contributions will not
exist at leading order of $\alpha_s$. We will come back to this issue later.
\par
\begin{figure}[hbt]
\begin{center}
\includegraphics[width=12cm]{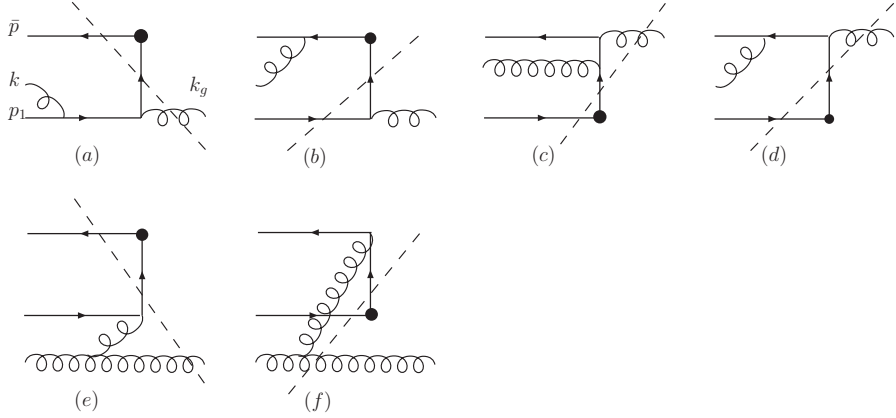}
\end{center}
\caption{Cut diagrams for the absorptive part of the amplitude
$P +\bar q \to \gamma^* + g$. The black dot is for the insertion of the electromagnetic
current, the broken line is for the cut.}
\label{Feynman-dg3}
\end{figure}
\par
We now study  the contributions from Fig.3. Actually, Fig.3a, Fig.3b and Fig.3e are also soft-pole contributions,
because they can be nonzero only if $k_g^- =0$. But it is not possible to have $k_g^-=0$ since the gluon
has a nonzero transverse momentum $\vec k_{g\perp} =- \vec q_\perp \neq 0$. So we have only three
diagrams to consider. It is straightforward to obtain the absorptive part from these diagrams.
By cutting the diagrams one should take care about that an extra minus sign is
associated with the insertion of the electromagnetic current.
\par
Taking the interference of Fig.3f with the complex conjugated Fig.2a as an example, denoted as
${\mathcal T}^{\mu\nu}\vert_{fa}$, which is the transition amplitude of
$ \bar q(\bar p) + q(p_1, \lambda_q) + g(k,\lambda_g) \to \bar q(\bar p) + q(p,\lambda_q)$,
we have
\begin{eqnarray}
{\mathcal T}^{\mu\nu}\vert_{fa} &=& -\frac{\pi g_s^3}{2N_c} \int \frac{d^4 k_g}{(2\pi)^4} (2\pi)\delta(k_g^2)
   \delta^4( p+\bar p -q- k_g) \delta ((\bar p+k -k_g)^2) \frac{1}{(k-k_g)^2 (p-k_g)^2}
\nonumber\\
   && \cdot \left \{ f^{abc} \bar u(p,\lambda_q) T^b \gamma^\alpha \gamma\cdot (p-k_g) \gamma^\nu \gamma\cdot \bar p
        \gamma^\rho T^c \gamma\cdot (k_g-k-\bar p) \gamma^\mu T^a u(p_1,\lambda_q)
\right.
\nonumber\\
   && \left. \cdot \left [ (k+k_g)_\rho \epsilon_\alpha (\lambda_g) +(-2k_g+k)\cdot \epsilon(\lambda_g)
      g_{\rho\alpha} + (k_g-2k)_\alpha \epsilon_\rho (\lambda_g) \right ] \right \},
\end{eqnarray}
where we take the spin- and color average of the initial antiquark.
Here we take the opportunity to correct a sign error in our previous work.
In \cite{MS1,MS2} the absorptive parts of amplitudes have been identified with a wrong sign.
Therefore, a minus sign should be added to the results for Sivers functions and structure functions
for SSA, where an absorptive part is involved. A minus sign should also be added to
the corresponding  formulas in the collinear factorization for SSA.
In the limit
$q^2_\perp \ll Q^2$,
the contribution comes from when the gluon in the intermediate state is almost collinear
to the quark, i.e., $k_g^+ \gg k_g^-$. Therefore we have for the on-shell condition for
the quark crossing the cut in Fig.3f:
\begin{equation}
\delta ((\bar p+k -k_g)^2) \approx \frac{1}{2 \bar p^-} \delta (k^+ -k_g^+), \ \ \ \ \
k_g^- =\frac{ q_\perp^2}{2 k_g^+}.
\end{equation}
Expanding ${\mathcal T}^{\mu\nu}\vert_{fa}$ in the limit $q^2_\perp \ll Q^2$  we obtain the leading contribution:
\begin{eqnarray}
{\mathcal T}^{\mu\nu}\vert_{fa} &=& g_\perp^{\mu\nu} \frac{g_s \alpha_s}{8\pi}  (N_c^2-1)
 \frac{1 }{(q^2_\perp)^2}
 (1-x_0) \sqrt{x_0}\delta(x-x_0) \delta (1-y) \left ( \frac{1}{1-x}\right )_+
\nonumber\\
   &&  \cdot \left [ i \vec q_\perp \cdot \vec \epsilon(\lambda_g) (1+x_0) -\lambda_q(1- x_0)
     \epsilon_\perp^{ij} \epsilon^i (\lambda_g) q_\perp^j \right ] + \cdots
\nonumber\\
    q^+ &=& xp^+, \ \ \ \ q^- =y \bar p^-.
\end{eqnarray}
The $\cdots$ in the second line  represents non-leading terms in $q_\perp$.
From this result one can derive the contribution from the interference between Fig.3f and Fig.2a
to $W^{\mu\nu}$ in the helicity basis as a $2\times 2$ matrix according to Eq.(7) at leading order
of $\alpha_s$ and in the limit $q^2_\perp \ll Q^2$:
\begin{eqnarray}
\nonumber\\
   W^{\mu\nu}\vert_{fa}     &=& g_\perp^{\mu\nu} \epsilon_{\perp}^{ij} \sigma_\perp^i q_\perp^j
   \frac{g_s \alpha_s}{8\pi} (N_c^2-1)
 \frac{1}{(q^2_\perp)^2}
 (1-x_0) \sqrt{2 x_0}\delta(x-x_0) \delta (1-y) \left ( \frac{1}{1-x}\right )_+
\nonumber\\
  &&  \cdot \left [ (1+x_0) -\vert \lambda_q \vert (1-x_0)\right ] + {\rm diagonal\ part}.
\end{eqnarray}
In the above we have separated the contributions from Eq.(25) to the tensor into two parts.
One is from the first term in Eq.(25), which does not depends on the spin of the quark
and is proportional to a trace of $\gamma$-matrices without $\gamma_5$.
Another part is from the second term and it depends on the quark spin with $\vert \lambda_q \vert =1$.
This part is proportional to a trace of $\gamma$-matrices with $\gamma^5 \gamma^-$.
We keep $\vert \lambda_q \vert$ to indicate the origin of the corresponding term.
The purpose of the separation will be discussed later.
One can extract $W_T^{(1)}$ from the above result.
For other diagrams we can perform the expansion in small $q_\perp$ as outlined in \cite{MS1}.
We find that the contributions from other three diagrams are power-suppressed by $q_\perp$
in comparison with $W^{\mu\nu}\vert_{fa}$. Therefore we have our final result
for the total $W_T^{(1)}$:
\begin{equation}
W_T^{(1)}(x,y,q_\perp)  =-\frac{g_s\alpha_s}{8\pi} (N_c^2-1)
 \frac{\sqrt{2 x_0}}{(q^2_\perp)^2}
 (1-x_0)\delta(x-x_0) \delta (1-y) \left [ (1+x_0) -\vert \lambda_q \vert (1-x_0)\right ]
  \left ( \frac{1}{1-x}\right )_+ .
\end{equation}
\par
If $W_T^{(1)}$ can be factorized with $T_F(x_1,x_2)$ alone as assumed in \cite{JQVY1}, with
our result of $T_F(x_1,x_2)$ in the last section and the leading order result
of the antiquark distribution
\begin{equation}
  \bar q (x) = \delta (1-x)
\end{equation}
we can derive the following  factorization formula:
\begin{eqnarray}
W_T^{(1)}(x,y,q_\perp) &=&
  \frac {\alpha_s}{ 2 \pi^2 (q^2_\perp)^2} \int_x^1 \frac{dy_1}{y_1}\int_y^1 \frac{ d y_2}{ y_2} \bar q(y_2)  T_F(x,y_1)
  H(x/y_1, y/y_2),
\nonumber\\
   H(\xi_1,\xi_2) &=& \delta (1-\xi_2) \frac{\xi_1}{(1-\xi_1)_+} + {\mathcal O}(\alpha_s).
\end{eqnarray}
This verifies the result
in \cite{MS1}, where the above factorization formula is verified with a finite quark mass
at leading order of $\alpha_s$. Beside the sign problem mentioned after Eq.(23), a slight modification to the factorization formula
in \cite{MS1} is introduced here. In \cite{MS1} the factor $\xi_1 =x /y_1$ in the numerator of the above
$H$ has been identified as $x$. This modification has no effect if we use our partonic
result to test the factorization, because after the integration over $y_1$ with $T_F$ given in
Eq.(13) one has $y_1=1$.
We note that the above result  of ${\mathcal H}$ is different than
that of the factorization formula derived formally in the collinear factorization approach in \cite{JQVY1}.
Using the factorization formula in \cite{JQVY1} instead of that in Eq.(29) and $T_F(x_1,x_2)$ in Eq.(13)
with the parton state, the partonic $W_T^{(1)}$ in Eq.(27) can not be reproduced.
\par
It has been shown in \cite{EKT,KVY} that $W_T^{(1)}$ should be factorized
not only with $T_F(x_1,x_2)$  but also with $\tilde T_F(x_1,x_2)$ defined in Eq.(14). These two quantities
characterize different contributions corresponding to different projections
of parton states. One involves with $\gamma^-$, another involves with $\gamma_5\gamma^-$.
The meaning of projections will be explained after Eq.(31).
The two different projections lead to two different contributions in $W_T^{(1)}$.
One contribution is given in our $W_T^{(1)}$ in Eq.(27) with the factor $(1+x_0)$ in $[\cdots]$.
This contribution should be factorized with $T_F$.
Another is given in Eq.(27) with $\vert \lambda_q \vert$  and should be factorized with $\tilde T_F$.
Therefore we derive the following factorized form with our partonic states:
\begin{eqnarray}
W_T^{(1)}(x,y,q_\perp) &=&
  \frac {\alpha_s}{  (2 \pi q^2_\perp)^2} \int_x^1 \frac{dy_1}{y_1}\int_y^1 \frac{ d y_2}{ y_2} \bar q(y_2)
  \left [  T_F(x,y_1)
  {\mathcal H}(x/y_1, y/y_2) + \tilde T_F(x,y_1) \tilde{\mathcal H}(x/y_1, y/y_2)  \right ]   ,
\nonumber\\
   {\mathcal H}(\xi_1,\xi_2) &=& \delta (1-\xi_2) \frac{1+\xi_1}{(1-\xi_1)_+} + {\mathcal O}(\alpha_s),
   \ \ \ \  \tilde {\mathcal H}(\xi_1,\xi_2) =\delta (1-\xi_2)+ {\mathcal O}(\alpha_s).
\end{eqnarray}
\par
Before ending this section we briefly discuss why the soft pole contribution appears
in the collinear factorization derived formally. In the formal approach
one works directly with the hadronic tensor in Eq.(18) and divide diagrams contributing
to the tensor into three parts: One part consists only of partons, and other two parts
can be defined as various parton density matrices of the two initial hadrons, respectively.
To illustrate the appearance of the soft pole contributions clearly, we take the light-cone
gauge $n\cdot G=0$. In the first step of the derivation one has the multi-parton density matrix
for the polarized hadron:
\begin{equation}
 \Gamma_{ji}(k_1,k_2) = \int\frac{ d^4 y_1 d^4 y_2}{(2\pi)^8}
   e^{ -iy_2 \cdot (k_2-k_1) -i y_1\cdot k_1 }
     \langle P_A,  s_\perp \vert
           \bar\psi_i (y_1 ) G^\mu(y_2) \psi_j(0) \vert P_A, s_\perp \rangle.
\end{equation}
The momentum $k_2-k_1$ is carried by the gluon entering the parton scattering.
Then one makes a collinear expansion for the density matrix:
\begin{eqnarray}
 \Gamma_{ji}(k_1,k_2) &=& \frac{1}{4} \left [ \gamma^- \right ]_{ij} \delta (k_1^-) \delta (k_2^-)
 \delta^2(\vec k_{1\perp}) \delta^2 (\vec k_{2\perp})
     \int\frac{ d y_1^- dy_2^-}{(2\pi)^2}
   e^{ -iy_2^-  (k_2^+-k_1^+) -i y_1^- k_1^+ }
\nonumber\\
    && \cdot  \langle P_A,  s_\perp \vert
           \bar\psi (y_1^- n ) \gamma ^+ G^\mu_\perp (y_2^- n ) \psi(0) \vert P_A, s_\perp \rangle
   +\cdots,
\end{eqnarray}
where we have given the projection with $\gamma^-$ explicitly. The $\cdots$
stand for another projection with $\gamma_5\gamma^-$ and other irrelevant terms.
It is clear from our calculation that the contribution with the projection
$\gamma_5\gamma^-$ corresponds to the second term in Eq.(25) or that with $\vert \lambda_q \vert$.
Assuming the above is nonzero, especially nonzero at $k_1^+ -k_2^+=0$,
and combining a factor $k_1^+ -k_2^+$ from the part of parton scattering, e.g., the factor $k^+=k_1^+ -k_2^+$
in the contribution
of Fig.1b, one can write the gauge field $G^\mu_\perp$ in the above as the field strength tensor $G^{+\mu}$.
At the end, one finds the soft pole contribution with  $k_1^+ -k_2^+=0$ or, e.g.,  $k^+=0$ in Fig.1b.
However, if we use our parton state to calculate the above density matrix, one finds that it is zero
at $k_1^+ -k_2^+=0$, as indicated by $T_F$ in Eq.(13) with $x_1=x_2$.
Our partonic $W_T^{(1)}$ at leading order of $\alpha_s$ does not contain any soft pole contribution. This is also
in the correspondence of $T_F(x,x)=0$ at the leading order. Therefore,
with our partonic results the soft-pole contributions in the collinear factorization derived
formally can not be identified.
Beside this, Our partonic $W_T^{(1)}$ can not be reproduced with our partonic $T_F(x_1,x_2)$ by using
the collinear factorization formula derived at hadron level.
\par
We have also performed calculations of the structure function $F_T^{(1)}$ for SSA in SIDIS, the same
factorization form is obtained by replacing corresponding quantities. This confirms our previous result
with a massive quark state in \cite{MS2}.

\par\vskip20pt
\noindent
{\bf 4. Sivers Function and TMD Factorization}
\par
In the TMD factorization the nonperturbative quantity responsible for SSA is the Sivers function.
In this section we perform a calculation with our partonic states to derive the TMD factorization
formula for SSA in Drell-Yan processes.
To define the Sivers function with QCD operators we introduce a gauge link
along the direction $u$ with $u^\mu =(u^+,u^-,0,0)$:
\begin{equation}
L_u (-\infty, z) = \left [ P \exp \left ( -i g_s \int_{-\infty}^0  d\lambda
     u\cdot G (\lambda u + z) \right ) \right ] ^\dagger .
\end{equation}
The Sivers function relevant for Drell-Yan process is defined in the limit $u^+ \ll u^-$:
\begin{eqnarray}
q_\perp (x,p_\perp)
\varepsilon_\perp^{\mu\nu} s_{\perp\mu}  p_{\perp\nu}
  & =& \frac{1}{4}  \int \frac{dz^- d^2 z_\perp}{(2\pi)^3}
                e ^{-i  xP^+ z^- + i \vec p_\perp \cdot \vec z_\perp }
\nonumber\\
    && \cdot  \left   \{
                \langle P, \vec  s_\perp \vert
                \bar\psi (z ) L_u ^\dagger (-\infty, z) \gamma^+
                L_u (-\infty, 0) \psi(0) \vert P,\vec s_\perp \rangle
             - (\vec s_\perp \to -\vec s_\perp ) \right  \} ,
\end{eqnarray}
with $z^\mu =(0,z^-,\vec z_\perp)$.
Beside the renormalization scale $\mu$, the Sivers function  also depends  on the parameter
$\zeta^2 =4 (u\cdot P)^2/u^2$. With the partonic state in Eq.(\ref{state})
we can calculate the function. As discussed
before, we will not consider the so called soft pole contribution because $k^+\neq 0$ in the partonic state.
At leading order we have contributions from the diagrams in Fig.4.
\par
\begin{figure}[hbt]
\begin{center}
\includegraphics[width=12cm]{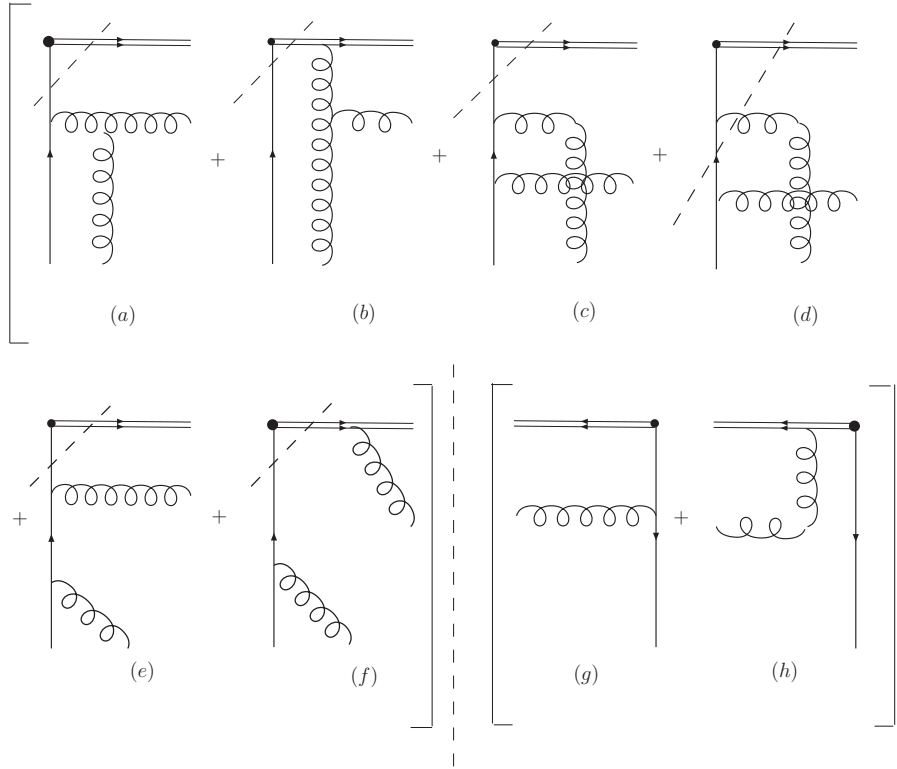}
\end{center}
\caption{The diagrams for the transition  amplitude
$  q(p_1, \lambda_q) + g(k,\lambda_g) \to q(p_1,\lambda_q)$
with the operator for Sivers function. the broken line is for the cut.}
\label{Feynman-dg3}
\end{figure}
\par
By taking the cut in Fig.4a to Fig.4f one should keep in mind that the gauge link represents a particle
moving from $t= -\infty$ and the energy flow along the gauge link is from the right to the left.
Also, an extra minus sign associated the joining vertex of the gauge link and the quark line
should be taken into account\cite{MS1}.
Although we have given possible cut for each diagram, but not every diagram can give
nonzero contribution. E.g., the cut in Fig.4f requires in the limit $u^+ \to 0$ that
the plus component of the momentum carried by the gluon in the intermediate state
must be zero. This is not possible for the finite $p_\perp$. Therefore, there is no contribution
from Fig.4f at the leading order. Similarly, we find that the contributions
from Fig.4a, Fig.4c, Fig.4d and Fig.4e are zero. The only nonzero contribution
comes from the interference of Fig.4b with Fig.4g. It is straightforward
to calculate
the non-diagonal part of the transition amplitude and to obtain Sivers function. We have:
\begin{equation}
 q_\perp (x,p_\perp) =
      -\frac{g_s\alpha_s}{4\pi (p_\perp^2)^2} N_c (N_c^2-1) x_0 \sqrt{2 x_0} \delta(x-x_0).
\end{equation}
Using the leading order result for the TMD parton distribution of the anitquark
\begin{equation}
   \bar q(z,k_\perp) =\delta(1-z) \delta^2 (\vec k_\perp)
\end{equation}
and the result $W_T^{(1)}$ given in Eq.(27), we can derive the TMD factorization for $W_T^{(1)}$ as
\begin{eqnarray}
W^{(1)}_T (z_1,z_2,q_\perp)
      &=&  \frac{1}{N_c}
  \int d^2 k_{1\perp}  d^2 k_{2\perp} \frac{\vec q_\perp \cdot \vec k_{1\perp}}{q^2_\perp}
  q_\perp (z_1, k_{1\perp}) \bar q(z_2, k_{2\perp})
    \delta^2 (\vec k_{1\perp} +\vec k_{2\perp} -\vec q_\perp ) H ,
\nonumber\\
   H&=& 1 + {\mathcal O}(\alpha_s).
\label{TMDFAC}
\end{eqnarray}
This result is in agreement with that verified by using massive parton state\cite{MS2} and
that derived formally. Beyond tree-level, a soft factor should be implemented for a complete
factorization.

\par\vskip20pt
\noindent
{\bf 5.  $T_F(x,x)$ and Its Relation to Sivers Function}
\par
Starting from the definitions of the twist-3 matrix element and Sivers function, one can formally derive a relation
between $T_F(x,x)$ and the first $k_\perp$-moment of $q_\perp(x,k_\perp)$ by invoking
discrete symmetries. The relation is\cite{Boer03,MW1}:
\begin{equation}
   T_F(x,x) = \int d^2 k_\perp k_\perp^2 q_\perp (x, k_\perp).
\end{equation}
In deriving the relation one assumes that $q_\perp (x, k_\perp)$ will decrease to $0$ rapidly
with $k_\perp \to \infty$. This assumption implies that one discards the U.V. subtraction
of the integral, i.e., the renormalization issue. Substituting our $q_\perp (x,k_\perp)$ in Eq.(35)
one sees clearly that the integral has an U.V. divergence.
In our previous work\cite{MS1} we have made an attempt
to modify the relation by taking the U.V. subtraction into account, where we have introduced
\begin{equation}
\tilde q_\perp (x, b) =  \int d^2 k_\perp e^{i \vec k_\perp \cdot \vec b} k^2_\perp q_\perp(x,k_\perp)
\end{equation}
to make the integral convergent in the large $k_\perp$-region. But with massive parton state
we failed to establish a relation between $\tilde q_\perp (x, b)$ and $T_F(x,x)$.
From previous studies of TMD factorization for unpolarized cases \cite{CS,CSS,JMY,JMYG}
it can be expected in general that certain perturbative relations exist between parton distributions
functions in collinear factorization and the corresponding parton distributions entering TMD
factorization.
The failure may indicate
certain problems when one uses massive parton states to study problems of SSA. In this section,
we will use our partonic state with massless partons to tackle the problem.
\par
We have already seen that $T_F(x,x)$ is zero at tree-level.
To obtain nonzero $T_F(x_1,x_2)$ at $x_1=x_2=x$, one has to consider it at the next-to-leading order.
At the next-to-leading order there are too many diagrams because of gauge links
between the operators in the definition of $T_F$. To reduce the number of diagrams we take
the light-cone gauge $n\cdot G=0$ so that all gauge links becomes unit.
In this gauge the gluon propagator reads:
\begin{equation}
\frac{-i}{q^2+ i\varepsilon} \left ( g^{\mu\nu} - \frac{q^\mu n^\nu + q^\nu n^\mu}{ n\cdot q}\right ).
\end{equation}
In the light-cone gauge one needs in principle to give a prescription for the $n\cdot q$ in the denominator
of the propagator. However,
it is irrelevant here, because we will not perform any loop integration in the $+$-component.
We will use the same partonic state as in Eq.(\ref{state}) to calculate the helicity flip amplitude
to extract $T_F(x,x)$.
In the light-cone gauge the leading contributions to $T_F(x,x)$ are given by the diagrams
in Fig.5. In the representation of contributions with diagrams, we do not take the representation
of cut diagrams. We take the representation with the product of operators in the definition
of $T_F$ as a T-order product. The two representations are equivalent\cite{Jaff}.

\begin{figure}[hbt]
\begin{center}
\includegraphics[width=12cm]{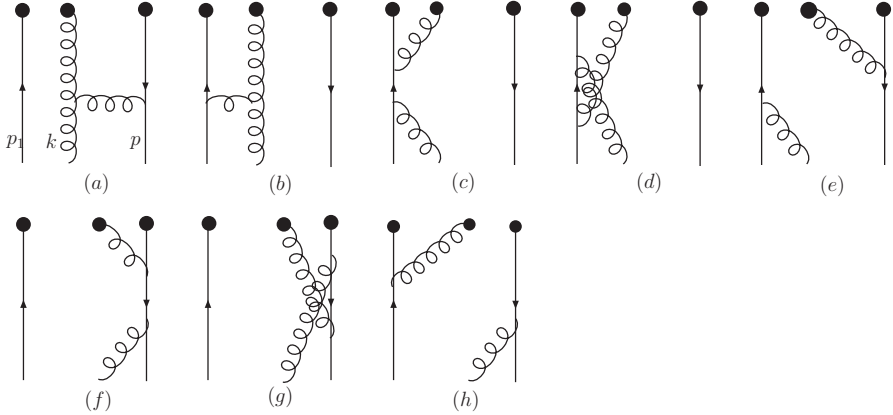}
\end{center}
\caption{The leading diagrams for $T_F (x,x)$ in the gauge $n\cdot G=0$.
The black dots represent inserting field operators
appearing in the definition of $T_F(x_1,x_2)$. }
\label{Feynman-dg3}
\end{figure}
\par
As mentioned in Section 3, there will be some problem if one deals diagrams in which two
collinear particles are combined into a divergent propagator. Here we encounter the same problem with
Fig.5c, Fig.5e, Fig.5f and Fig.5h. We will give the quark a small mass $m$ to regularize the divergent
propagator and set $m=0$ in our final result. In fact, these diagrams do not bring any trouble as we will see.
\par
We first consider the contribution from Fig.5b. For $T_F(x_1,x_2)$ with $x_1 =x_2$ we find that its
contribution is proportional to the integral
\begin{equation}
\int\frac{ dq^-}{2\pi} \frac{ 1}{(p_1 -q)^2 +i\varepsilon} \cdot \frac{1}{ q^2+i\varepsilon} =0
\end{equation}
where $q$ is the momentum of the gluon exchanged between the quark- and gluon line.
$q^+$ is fixed as $q^+ = -k^+$ because $x_1 =x_2$. With the fixed $q^+$ one finds that the above integral
is zero. Hence Fig.5b will not contribute to $T_F(x,x)$. Because of the same reason,
Fig.5d and Fig.5g will also give no contributions to $T_F(x,x)$. For Fig.5e
with arbitrary $x_1$ and $x_2$ the contribution is proportional to the integral
\begin{equation}
\int\frac{d^2 q_\perp }{(2\pi)^2} \frac{q^\mu_\perp}{q^2+i\varepsilon} \cdot \frac{1}{(p-q)^2+i\varepsilon} =0.
\end{equation}
The integral is zero because of the rotation covariance in the transverse space.
With the similar reason one finds that Fig.5c, Fig.5f and Fig.5h
do not contribute to $T_F(x_1,x_2)$. Therefore, only Fig.5a gives
nonzero contribution to $T_F(x,x)$. In the calculation we will have U.V.- and collinear divergence.
We use dimensional regularization by taking the dimension of the transverse space as $2-\epsilon$
to regularize the divergences. We obtain then:
\begin{equation}
T_F(x,x,\mu) = -\frac{g_s\alpha_s}{4} N_c (N_c^2-1) x_0 \sqrt{2x_0} \delta(x-x_0)
 \left [ -\left ( \frac{2}{\epsilon} -\gamma +\ln 4\pi \right ) + \ln\frac{\mu^2}{\mu_c^2} \right ].
\end{equation}
In the above we have subtracted the U.V. pole and introduced the renormalization scale $\mu$.
The remaining pole is a collinear divergence with its corresponding scale $\mu_c$.
\par
Taking our result of $q_\perp(x,k_\perp)$ in the last section we obtain:
\begin{eqnarray}
\tilde q_\perp (x, b)
    = -\frac{g_s\alpha_s}{4} N_c (N_c^2-1) x_0 \sqrt{2 x_0} \delta(x-x_0)
       \left [ -\left ( \frac{2}{\epsilon} -\gamma +\ln 4\pi \right ) -\ln \frac{ b^2 e^{2\gamma} \mu_c^2 }{4} \right ],
\end{eqnarray}
where the pole is for a collinear divergence. Comparing the above results, we find
that $T_F(x,x)$ and $\tilde q_\perp(x,b)$ contain the same long distance effect represented by the
collinear divergence. Hence a relation between them can be established. With our results we get:
\begin{equation}
  T_F(x,x,\mu) \biggr\vert_{\mu =\frac{2 e^{-\gamma}}{b}} = \tilde q_\perp (x, b) + {\mathcal O}(\alpha_s).
\end{equation}
This result is in certain sense of factorization, where $T_F(x,x)$ and $\tilde q_\perp(x,b)$
have the same nonperturbative effects represented by the collinear divergence, and the difference is a perturbative effect. It will be interesting
to study the relation at higher orders of $\alpha_s$.
\par\vskip20pt
\noindent
{\bf 6. Summary}
\par
There are two factorization approaches for SSA. The factorization formulas in different approaches
are formally derived with the diagram expansion at hadron level, where one separates diagrams involving hadrons
into a parton-scattering part and hadronic parts. The hadronic parts involve hadronic states and
are characterized by various parton density matrices of hadrons.
If these factorization hold, they also hold when one replaces hadronic states with partonic states,
because any proved factorization is a general property of QCD's Green functions and
the perturbatively calculable parts do not depend on hadrons.
On the other hand, it is also important to examine any factorization with partonic states
and eventually to prove the factorization.
It is a nontrivial task to examine factorizations for SSA at parton level, because
one must have helicity-flip and nonzero absorptive parts of scattering amplitudes
to generate SSA and the helicity of a massless quark
in QCD is conserved.
\par
With the motivation discussed in detail in the introduction,
we have made a first step to examine these factorization approaches for SSA in \cite{MS1,MS2},
where we have used single massive-quark state to replace the transversely-polarized hadron.
The nonzero mass is responsible for the needed helicity-flip. By examining all elements
in the factorizations we have found that the formally derived TMD factorization
holds at our parton level, while the formally derived collinear factorization
has the perturbative coefficient function different than that derived from our partonic results.
\par
In this work, we construct a simple- and complicated parton states which consists massless partons and
are suitable for studying factorization approaches for SSA. With the constructed
partonic states we verify all factorization results in \cite{MS1,MS2} derived
with a single massive-parton state. The results obtained with the simple parton state
are the same as those with the complicated parton state.
Because partons here are massless, our work
gives an example how the correlation between spins of parton can change
the helicity of a state. With our results we have established the relation
between the twist-3 matrix element $T_F(x,x)$ and the first $k_\perp$-moment of Sivers function
by taking U.V. subtraction into account.
\par
With our results presented here, we have found that the perturbative coefficient in the collinear
factorization of SSA is different that derived in a formal way, i.e., with the
diagram expansion involving hadrons. Some reasons for the difference
have been discussed in Sec. 3..
The difference indicates that a further study of the collinear
factorization for SSA is needed.
\par
Having examined the two factorization approaches with the constructed parton states
at leading but nontrivial order of $\alpha_s$, one can study higher order corrections
in the two factorizations. We notice that recently parts of evolutions
of twist-3 matrix elements have been derived\cite{KQiu,ZYL,BMPEV}. With our partonic state
it is straightforward to derive complete evolutions of twist-3 matrix elements.
It is interesting to note that there is a discrepancy between the result of \cite{KQiu,ZYL} and
that of \cite{BMPEV}. A part of reasons for this discrepancy is that the contributions to SSA and
to the evolutions from the forward scattering $q\bar q \to g$ and the reversed scattering have been
not considered so far in the collinear factorization. Following our work one can also construct
a more suitable parton state to identify these contributions. A study
for complete contributions in the collinear factorization of SSA
is under the way.

\par\vskip20pt
\par\noindent
{\bf\large Acknowledgments}
\par
We thank Prof. Y. Liao for an interesting discussion.
This work is supported by National Nature Science Foundation of P.R. China((No. 10721063, 10575126 and 10975169).
\par

\end{document}